\documentstyle[11pt]{article}
\begin{document}
\baselineskip 18pt
\begin{titlepage}
\centerline{\large\bf
           Estimate the Ranges of $\rho$ and $\eta$}
\centerline{\large\bf
           only from the Kobayashi-Maskawa Matrix Elements }
\centerline{\large\bf
       of the First Two Generations}
\vspace{1cm}

\centerline{ Yong Liu }
\vspace{0.5cm}
{\small
\centerline{\bf  Lab of Numerical Study for Heliospheric Physics}
\centerline{\bf  Chinese Academy of Sciences,
                 P. O. Box 8701, Beijing 100080, P.R.China}
}
\vspace{2cm}

\centerline{\bf Abstract}
\vspace{0.3cm}

Based on the relation between weak $CP$ phase and the other three
mixing angles in Cabibbo-Kobayashi-Maskawa (CKM) matrix postulated
by us before, the ranges of $\rho$ and $\eta$ have
been estimated by using the best known two KM matrix elements $V_{ud}$
and $V_{cd}$ (or $V_{us}$). It is found that, the upper limit on
$\eta$ is about 0.008 which is consistent with that estimated
by Wolfenstein more than ten years ago but far small than the present
popular estimation.\\ \\
PACS number(s): 12.10.Ck, 13.25.+m, 11.30.Er

\vspace{5cm}
\noindent
Email address: yongliu@ns.lhp.ac.cn
\end{titlepage}

\centerline{\large\bf
           Estimate the Ranges of $\rho$ and $\eta$}
\centerline{\large\bf
           only from the Kobayashi-Maskawa Matrix Elements }
\centerline{\large\bf
       of the First Two Generations}

\vspace{1cm}

Quark mixing and $CP$ violation is one of the most interesting
and important problem in weak interaction [1-4]. In the Minimal Standard
Model (MSM), They are described by the unitary Cabibbo-Kabayashi-Maskawa
(CKM) matrix, which takes the following form [5-7]
\begin{equation}
V_{KM}= \left (
\begin{array}{ccc}
   c_1 & -s_1c_3& -s_1s_3 \\
   s_1c_2 & c_1c_2c_3-s_2s_3e^{i\delta}& c_1c_2s_3+s_2c_3e^{i\delta}\\
   s_1s_2 & c_1s_2c_3+c_2s_3e^{i\delta}& c_1s_2s_3-c_2c_3e^{i\delta}
\end{array}
\right )
\end{equation}
with the standard notations $s_i=\sin\theta_i$ and $c_i=\cos\theta_i$
being used. In fact, we always can take $0 < \theta_i < \frac{\pi}{2} $
and $- \pi < \delta < \pi$ by a suitable choice of phase.

To make it convenient to use the CKM matrix in the actual calculation,
Wolfenstein parametrized it as [8]
\begin{equation}
V_{W}= \left (
\begin{array}{ccc}
   1-\frac{1}{2}\lambda^2 & \lambda & A\lambda^3(\rho-
   i\eta+i\eta\frac{1}{2}\lambda^2) \\
   -\lambda & 1-\frac{1}{2}\lambda^2-i\eta A^2 \lambda^4 &
   A\lambda^2(1+i\eta\lambda^2)\\
   A\lambda^3(1-\rho-i\eta) & -A\lambda^2 & 1
\end{array}
\right ).
\end{equation}

Actually, one can take different parametrization [7][9-13] in
different cases. They are only for the convenience in discussing
the different concrete question, but the physics does not change when
adopting various parametrizations.

In Eq.(2), $\lambda$ and $A$ are the two better known parameters. But,
due to the uncertainty of hadronic matrix elements and other
reasons, we can not extract more information about $\rho$ and $\eta$
from experimental results. Up to now, we still know little about
them. More than ten years ago, Wolfenstein estimated
that the upper limit on $\eta$ is about $0.1$ [8], but the recent
estimate on $\rho$ is about $0$ and $\eta$ about $0.35$ [9][14].

The center purpose of this short letter
is to give a limit on the ranges of
$\rho$, $\eta$ and their dependence on each other by only
use of the two best known KM matrix elements $V_{ud}$ and $V_{cd}$ or
$V_{us}$.
According to the
usual view point, this is impossible. But, by using the constraint on
weak $CP$ phase and the three mixing angles postulated by us before,
we can do so.

In Ref.[15], we have found that the weak $CP$ phase and the other
three mixing angles satisfy the following relation 
\begin{equation}
sin\frac{\delta}{2}=\sqrt{\frac{sin^2\theta_1+sin^2\theta_2+sin^2\theta_3
-2(1-cos\theta_1 cos\theta_2 cos\theta_3)}{2 (1+cos\theta_1) (
1+cos\theta_2) (1+cos\theta_3)}}
\end{equation}
where $\theta_i\; (i=1,\; 2,\; 3)$ are the corresponding angles in the
standard KM parametrization matrix Eq.(1).
Here, we have taken $\delta$ presented in Eq.(1) as a certain geometry
phase. In fact, the geometry meaning of Eq.(3) is as follows

$\delta$ is the solid angle enclosed by $\theta_1$, $\theta_2$ and
$\theta_3$, or the area to which the solid angle corresponds on a unit
sphere.

To make $\theta_1$, $\theta_2$ and $\theta_3$ enclose a solid
angle, the condition
\begin{equation}
\theta_i+\theta_j > \theta_k  \;\;\;\; (i\neq j \neq k \neq i.
\;\; i,j,k=1,2,3)
\end{equation}
must be satisfied. 

Based on Eq.(3) and Eq.(4), we can proceed our discussion.

From the two best known matrix elements
\begin{equation}
V_{ud}=c_1
\end{equation}
and
\begin{equation}
V_{cd}=s_1 c_2
\end{equation}
in KM parametrization, we can get $\theta_1$ and $\theta_2$. According
to Eq.(4), we have
\begin{equation}
\mid \theta_1-\theta_2 \mid < \theta_3 < \theta_1+\theta_2.
\end{equation}
When we let $\theta_3$ vary in the above domain, we can obtain a permitted
domain of $\delta$ from Eq.(3). Considering the relation between KM
parametrization and Wolfenstein's parametrization [16]
\begin{equation}
\rho=\frac{\sin \theta_3}{(\sin^2\theta_2+\sin^2\theta_3+
           2 \sin\theta_2\sin\theta_3\cos\delta)^{\frac{1}{2}}}
\end{equation}
\begin{equation}
\eta=\frac{\sin\theta_2 \sin\theta_3 \sin\delta}{(\sin^2\theta_2+
           \sin^2\theta_3+2 \sin\theta_2\sin\theta_3\cos\delta)}
\end{equation}
taking use of Eq.(8) and Eq.(9), we can finally
get the ranges of $\rho$, $\eta$ and their dependence on each other.

Take the experimental values [7]
\begin{equation}
V_{ud}=0.975\pm0.001 \;\;\;\;
V_{cd}=0.2205\pm0.0018
\end{equation}
as input, when fixing $V_{ud}$ and $V_{cd}$ at their central values,
the dependence of $\eta$ on $\rho$ is shown in Fig.(1). Taking
the errors of $V_{ud}$ and $V_{cd}$ into account, we can get the permitted
ranges of $\rho$ and $\eta$, the result is given in Fig.(2).

From Fig.(2), we find that
a relative large $\eta$ corresponds to a relative
narrow window of $\rho$ around $0.55$.

In conclusion, we find that, the upper limit on $\eta$ is about 0.008,
this is consistent with the prediction given by
Wolfenstein fifteen years ago
[8], but very lower than the present estimate accepted by most
people. According to Wolfenstein [8][17], the smaller
$\frac{\mid \epsilon' \mid}{\mid \epsilon \mid} $,
the smaller $\eta$, so, if we can see a small
$\frac{\mid \epsilon' \mid}{ \mid \epsilon \mid} $ in the
future experiments, especially,
if $\frac{\mid \epsilon' \mid}{ \mid \epsilon \mid} $  can
be small to the order of about
$10^{-5}$ or less, we will win a strong support to our viewpoint.

\end{document}